# On the viability of some Emergent Universe models

S. Ghose
Department of Physics, University of North Bengal
West Bengal, India – 740013
e-mail: *souviknbu@rediffmail.com*
B. C. Paul
Department of Physics, University of North Bengal
West Bengal, India – 740013
e-mail: *bcpaul@iucaa.ernet.in*

**Abstract:** A particular class of flat Emergent Universe scenario is studied in light of recent observational data. Observationally permissible ranges of values are obtained for the model parameters. The class of model studied here can accommodate different composition of matter-energy as cosmic fluid. It is found that recent observations favour some compositions over others while some compositions can be ruled out with some level of confidence.

**Introduction:** It is well accepted today that we are living in an expanding universe. The overwhelming success of standard Big-Bang cosmology over any other model also suggests that our universe began at some finite past. Apart from having a time-like singularity in the past, where the laws of physics seem to break down there are other issues related to very early universe which require a phase of rapid expansion, commonly known as 'Inflation' [7]. Recent observations also suggest that our universe is now passing through a phase of acceleration. One needs to include a cosmological constant ($\Lambda$) in order to explain this phase within the standard model of cosmology. Unfortunately the physics of the inflation and cosmological constant, which needs to be of an unusually small value, is not well understood. This is where other models make their way into the literature. If possible one would like to have a cosmological model which can avoid initial singularity and accommodate late time acceleration. Emergent Universe (EU) models are studied in hope that an ever existing model might resolve the issues regarding inflation at the same time naturally escaping the messy situation of initial singularity [4,5,8]. Many such models, based on either flat or closed universe, has been proposed in literature in the frame work of Brane world gravity [1], Gauss-Bonnet gravity [11], Brans-Dicke theory [2] etc. One interesting model was proposed by Mukherjee *et.al.* [9] in the frame work of general relativity which would accommodate a late time acceleration as well. Although they found an exact solution of Friedmann equations in flat universe they had to compromise with the equation of state which they considered as:

$$p = A\rho - B\rho^{1/2} \qquad (1)$$

This type of equation of state (EOS) is a special form of a more general polytrophic EOS :

$$p = A\rho - B\rho^\alpha \tag{2}$$

A string inspired phenomenology can be found for such EOSs and most of the time the models interpolate between a matter (or radiation) dominated era and a late time de Sitter phase. Recently Paul, Thakur and Ghose [10] studied the model in view of observational data and obtained best fit values for the EOS parameters, i.e., 'A' and 'B'. It was shown that observations favour a small but negative value of 'A' but 95% confidence level can allow up to a small but positive value for 'A'. This analysis was how ever done considering a best fit value for 'K', another parameter appearing in the original theory [9]. It is worth to note that in the original model 'K' creeps in as an integration constant and can thus be fixed to some other value by adjusting initial conditions which were quite arbitrary. Later in another work Paul, Ghose and Thakur [11] considered this as a possibility and analyzed the viability of a model with 'A=0'. The importance of choosing a particular value for 'A' lies in the fact that in this model it decides the composition of matter energy content of the universe. In this paper we analyze the model for other possible values of 'A' which might suggest a very different matter energy content of the universe and make a comparative study of the models.

**Method and Field equations:** Friedmann equation for a flat universe reads as :

$$H^2 = \left(\frac{\dot{a}}{a}\right)^2 = \frac{\rho}{3}$$

Also, the conventional conservation equation holds :

$$\frac{d\rho}{dt} + 3H(\rho + p) = 0 \tag{4}$$

Along with the EOS given by eq. (1) the conservation equation yields[9]:

$$\rho(z) = \left(\frac{B}{A+1}\right)^2 + \frac{2BK}{(A+1)^2}(1+z)^{\frac{3(A+1)}{2}} + \left(\frac{K}{A+1}\right)^2 (1+z)^{3(A+1)} \tag{5}$$

where 'z' is the redshift. One can define Hubble parameter 'H' as : $H = \frac{\dot{a}}{a} = -\frac{1}{1+z}\frac{dz}{dt}$. The first term in the right hand side of eq. (5) is a constant and one might be tempted to interpret it as the dark energy (or cosmological constant) component. Eq. (5) can be written in a simpler form:

$$\rho(z) = \rho_1 + \rho_2 (1+z)^{\frac{3(A+1)}{2}} + \rho_3 (1+z)^{3(A+1)} \tag{6}$$

Here $\rho_1, \rho_2, \rho_3$ denote different components of matter-energy density at present epoch. The Friedmann equation can be re-written in terms of density parameters ($\Omega$):

$$H^2(z) = H_0^2 \left( \Omega_{Const} + \Omega_1 (1+z)^{\frac{3(A+1)}{2}} + \Omega_2 (1+z)^{3(A+1)} \right) \quad (7)$$

Here the definition of density parameter is usual and obvious ($\Omega = 8\pi G\rho / 3H_0^2$). Note that once we make a choice for the parameter 'A' it fixes the way of evolution for the two variable components of , thereby specifying two components of cosmic fluid other than the constant one. The present values of the density parameters then depend on the value of the other two parameters, namely 'B' and 'K'. Consequently, after we have assigned 'A (say $A_0$)' a value of our choice we can write:

$$H^2(H_0, B, K, z) = H_0^2 E^2(B, K, z) \quad (8)$$

with,

$$E^2(B, K, z) = \Omega_{Const} + \Omega_1 (1+z)^{\frac{3(A_0+1)}{2}} + \Omega_2 (1+z)^{3(A_0+1)} \quad (9)$$

**Analysis with OHD data:** For our first analysis we use a set of measurement for Hubble parameter at different redshifts. More specifically we use twelve data points as given in Observed Hubble Data (OHD) [13]. For this we define a chi square function:

$$\chi^2_{OHD} = \sum \frac{\left( H^2_{Theory}(H_0, B, K, z) - H^2_{Obs} \right)}{2\sigma_z^2} \quad (10)$$

where $H_{Theory}$ and $H_{Obs}$ are respectively the theoretically predicted value and the observed value of the Hubble parameter at a given redshift and $\sigma$ is the corresponding error in measurement. The parameter $H_0$ is a nuisance parameter here and we marginalize over it considering a fixed prior. We consider $H_0=72(8)$. The reduced chi square function is given as:

$$\bar{\chi}^2_{OHD} = -2\ln \left[ \int e^{-\frac{\chi^2_{OHD}}{2}} P(H_0) dH_0 \right] \quad (11)$$

Where $P(H_0)$ is the prior distribution function. The 66% (solid), 95% (dashed) and 99% (dotted) confidence contours on B-K plane, obtained for different values of 'A', are shown in fig. 1(a, b, c) in Appendix. Best fit values are listed in Table.1. For the case 'A=0' please refer to [11].

**Analysis with a BAO peak parameter:** It is only reasonable that the analysis is cross checked with some observation which is independent of measurement of $H_0$ and thus does not depend on any particular dark energy model either. For this we follow a method as proposed by Eisenstein *et. al.*[3]. For low redshift measurements in a flat universe a model independent BAO (Baryon Acoustic Oscillation) peak parameter can be defined as :

$$\Re = \frac{\sqrt{\Omega_m}}{E(z_1)} \left( \frac{\int_0^{z_1} \frac{dz}{E(z)}}{z_1} \right)^{2/3} \quad (12)$$

where $\Omega_m$ is the total density parameter for the matter content of the universe. For a detailed description of the above defined parameter and the approximations involved the reader is referred to [3]. We can now define a chi square function :

$$\chi^2_{Tot} = \overline{\chi}^2_{OHD} + \chi^2_{BAO} \quad \text{where,} \quad \chi^2_{BAO} = \sum \frac{(\Re - 0.469)^2}{(0.017)^2} \quad (13)$$

Here we have used the observed value of $\Re = 0.469(0.017)$ as measured in [3] from SDSS data for LRG (luminous red galaxies). The contour diagrams are shown in fig. 2(a, b, c). and the best fit values are listed in Table 1.

| A | OHD | | OHD+BAO | | $\chi^2$ d.o.f (tot) | $\Omega_1$ | $\Omega_2$ | $\Omega_\Lambda$ |
|---|---|---|---|---|---|---|---|---|
|   | B | K | B | K |   |   |   |   |
| -1/3 | 0.3048 | 0.4620 | 0.2107 | 0.5600 | 1.4310 | 1.4551 | -1.1369 | 0.6872 |
| 1/3  | 0.2940 | 0.4281 | 0.5093 | 0.3032 | 1.3037 | 0.2874 | 0.0917  | 0.6209 |
| 1    | 0.3541 | 0.3596 | 0.8363 | 0.1252 | 1.8888 | 0.4408 | 0.0144  | 0.5448 |

**Table-1**

**Analysis of Density Parameters:** With the best fitted values for 'B' and 'K', as obtained from the joint analysis of OHD and BAO data, we plot 66% (solid), 95% (dashed) and 99% (dotted) confidence contours on $\Omega_1$-$\Omega_2$ plane (fig. 3a, 3b and 3c in Appendix) and obtain the best fit values of density parameters for different values of 'A' (Table-1). It is easily seen that the values obtained for 'A=-1/3' are unrealistic.

**Findings and Remarks:** In the original work by Mukherjee *et. al.*[9] it has been shown that for 'A=-1/3' one obtains a universe with dark energy, domain wall and cosmic string; for 'A=1/3' the universe contains dark energy cosmic string and radiation while for 'A=1' the cosmic fluid behaves like a composition of dark energy dust and stiff matter.

We can compare our results, depicted in Table-1, with that obtained for 'A=0' in [11]. It seems that although recent observational data seem to favour the model with 'A=0' and can rule out the 'A=-1/3' within 95% confidence from the unrealistic present day values of DP. Within the realm of our present analysis though 'A=1 and 1/3' models survive within good confidence. However, a better analysis perhaps can be done with SNIa data which is more available recently that OHD data. Till today there is no satisfactory theory of structure formation in EU which could have been very useful in obtaining observational bounds from dynamical considerations. With these analyses one might obtain more stringent constraints on the model parameters and single out a specific EU model. These scopes will be considered in future works.

**Acknowledgement**: SG thanks Dr B. C. Paul for valuable discussion and CSIR for providing Senior Research Fellowship.

# Appendix

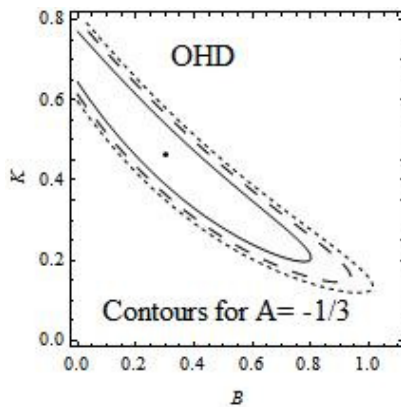
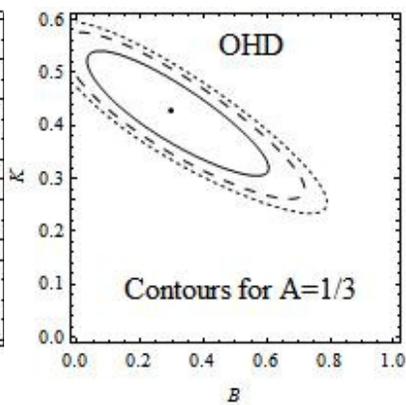
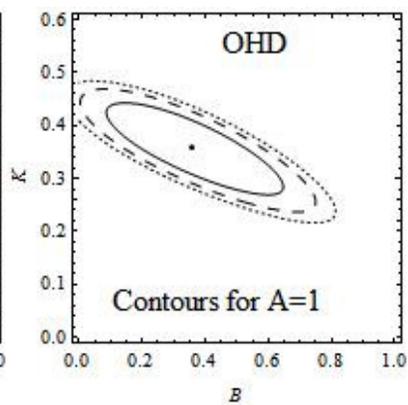

Fig. 1-a    Fig. 1-b    Fig. 1-c

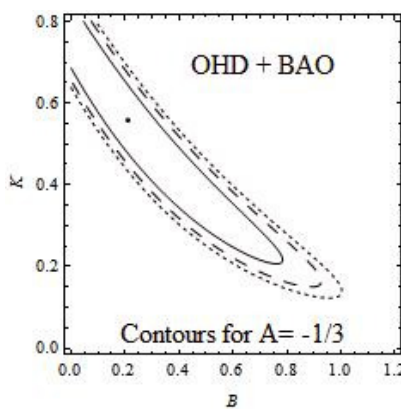
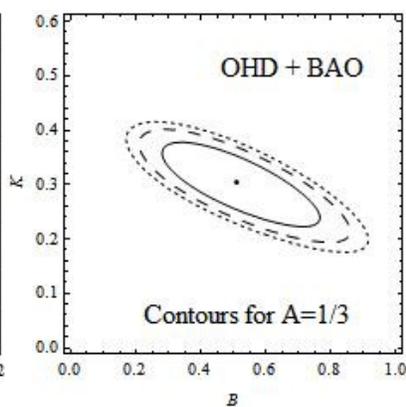
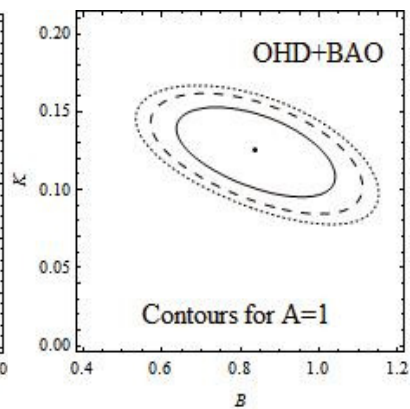

Fig. 2-a    Fig. 2-b    Fig. 2-c

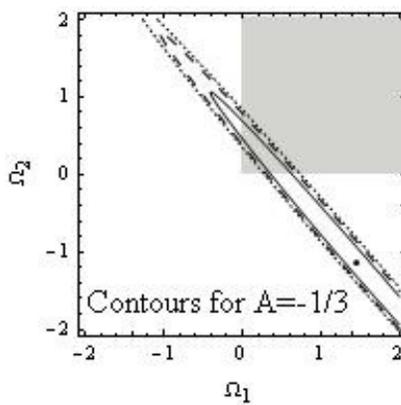
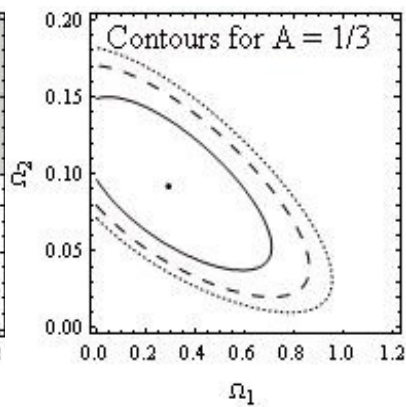
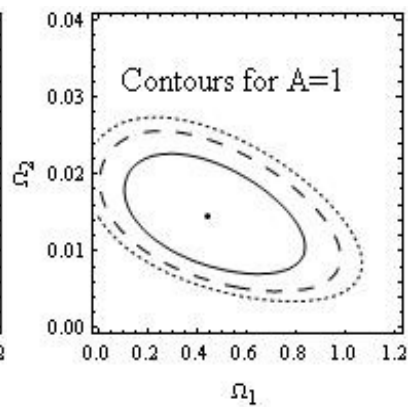

Fig. 3-a    Fig. 3-b    Fig. 3-c